\documentclass[aps,twocolumn,nopacs,superscriptaddress]{revtex4}
\usepackage{amsmath,amssymb,graphicx}

\begin{document}

\title{Non-inertial lateral migration of vesicles in bounded Poiseuille flow}

\author{Gwennou Coupier}\email[]{gcoupier@spectro.ujf-grenoble.fr}\affiliation{Laboratoire de Spectrom\'{e}trie Physique, CNRS - UMR 5588,
Universit\'{e} Grenoble I, B.P. 87, 38402 St Martin d'H\`{e}res Cedex,
France}
\author{Badr Kaoui}\affiliation{Laboratoire de Spectrom\'{e}trie Physique, CNRS - UMR 5588,
Universit\'{e} Grenoble I, B.P. 87, 38402 St Martin d'H\`{e}res Cedex,
France}
 \affiliation{Universit\'{e} Hassan II - Mohammedia, Facult\'{e} des
Sciences Ben M'Sik, Laboratoire de Physique de la Mati\`{e}re Condens\'{e}e,
BP 7955 Casablanca, Morocco}
\author{Thomas Podgorski}
\author{Chaouqi Misbah}

\affiliation{Laboratoire de Spectrom\'{e}trie Physique, CNRS - UMR 5588,
Universit\'{e} Grenoble I, B.P. 87, 38402 St Martin d'H\`{e}res Cedex,
France}

\date{\today}

\begin{abstract}Cross-streamline non-inertial migration of a vesicle in a bounded Poiseuille flow is investigated
experimentally and numerically. The combined effects of the walls and of the curvature of the velocity profile induce a movement towards the center
of the channel. A  migration law (as a function of relevant
structural and flow parameters) is proposed that is consistent with experimental and numerical results. This similarity law
markedly differs from its analogue in unbounded geometry.  The
dependency on the reduced volume $\nu$ and viscosity ratio
$\lambda$ is also discussed. In particular, the
migration velocity becomes non monotonous as a function of $\nu$
beyond a certain $\lambda$.
\end{abstract}

\maketitle

Flow of confined soft entities, such as vesicles (closed quasi-inextensible lipid membranes), or blood cells in the circulatory system
and in microfluidic devices, is a problem of a paramount importance
with both fundamental and practical interests. While inertial effects can induce lateral migration of any flowing body in a channel~\cite{inertial1,inertial2,inertial3}, the ability of these soft
entities to adapt their shapes under non-equilibrium conditions
gives them the possibility to migrate transversally even at low Reynolds number. Transverse migrations induce non uniform lateral distributions of
the suspended entities, which has important consequences on the
rheology of a confined suspension (e.g. the Fahraeus-Lindquist
effect in blood vessels~\cite{fung}), or should
impact on transport  efficiency in the various sorting microfluidic devices that
are now being developed~\cite{pamme07}.

Despite the considerable interest for transverse migration in many
circumstances, there is, to our knowledge, yet no quantitative law
that would allow one to relate the lateral migration velocity of a
deformable entity flowing in a channel, even isolated, with its
position, its mechanical properties, and the flow parameters. We propose a law for the case of a single
vesicle placed in a bounded Poiseuille flow, from experiments in microfluidic devices as well as simulations based on
the boundary integral method.

The behavior of vesicles under unbounded shear flow has been the
subject of several theoretical \cite{keller82,beaucourt04,vestheo1,vestheo2,vestheo3}
and experimental \cite{dehaas97,mader06,kantsler0506} studies. When
the viscosity ratio between the inner and the outer fluids is small,
vesicles perform a tank treading dynamics where the orientation of
the main axis of the vesicle is constant and the membrane undergoes
a tank treading motion. In a bounded Poiseuille flow, tank-treading vesicles experience a transverse force and reach the center where they assume a steady shape. The latter stage  has been described in several
papers~\cite{vescentre1,vescentre2,vescentre3}. Migration in Poiseuille flows has been also reported on
capsules~\cite{risso06,bagchi07,secomb07}, red blood
cells~\cite{bagchi07,secomb07}, and
drops~\cite{mortazavi00,griggs07}. The latter seem to have a very different behavior: depending on the viscosity ratio and the confinement,  the reported equilibrium positions are not always on the centerline. In addition, a drop interface and a vesicle membrane are mechanically different: a vesicle has a constant surface area and its equilibrium shape is not a sphere in general. As a consequence of these intrinsic differences the vesicle shape under flow is described (in the co-moving frame), to leading order, by a nonlinear shape equation while a drop obeys a linear equation~\cite{vestheo1}.

Lateral migration has two distinct sources. (i)  A wall-induced lift
force~\cite{olla97,cantat99, sukumaran01,abkarian02}. In agreement
with the numerical and theoretical studies for simple shear flows~\cite{olla97,sukumaran01}, we showed recently that the
migration velocity decreases like $1/y^2$, where $y$ is the distance
to the wall of the center of mass of the vesicle~\cite{callens08}.
(ii) The non-constant shear rate in a parabolic velocity profile (even unbounded) leads to a
subtle interplay between the gradient of shear and the
shape~\cite{kaoui08}, resulting in migration towards the center with
a constant drift velocity except near the center-line. In a realistic channel, both effects coexist and we shall see that
this leads to a new and nontrivial non-inertial migration law. 

The considered microfluidic channel is straight and has a rectangular cross section. The flow direction is $Ox$, and
we investigate lateral migration  along $Oy$. Let $2w$ denote the
channel width in the $y$-direction, and $v_0$ the imposed flow
velocity at the center of the channel in the absence of vesicle.
 The two walls are located at $y=0$ and $y=2w$.

A vesicle is characterized by two geometrical parameters: its
effective radius $R_0$, determined from its constant volume $\mathcal{V}$ by
$R_0=(3 \mathcal{V} / 4 \pi)^{1/3}$, and its reduced volume
$\nu=\mathcal{V}/\big[4\pi(\mathcal{S}/4\pi)^{3/2}/3\big]$
($\mathcal{S}$ is the constant area of the vesicle) characterizing vesicle
deflation. Volumes are calculated  at lift-off (see below) by assuming axisymmetric shape
about the vesicle's  main axis. The viscosity ratio is defined as
$\lambda=\eta_{in}/\eta_{out}$, where  $\eta_{in},\eta_{out}$ denote
the inner and the outer viscosities. Relevant space and time scales are the vesicle radius $R_0$ and the characteristic time needed by the vesicle to relax to its equilibrium shape (in the absence of imposed flow), which is given by $\tau=\eta_{out} R_0^3/\kappa$, where $\kappa \sim 20 k_B T$ is the membrane's bending rigidity (typically, $\tau \sim 10$ s). 

Therefore dynamics depends a priori on the four dimensionless parameters ($\hat{w}\equiv w/R_0$,
$\hat{v}_0\equiv v_0 \tau/R_0$, $\nu$, $\lambda$). We consider first the case $\lambda\simeq
1$. The strategy is to
vary $\hat{w}$ and the imposed velocity $\hat{v}_{0}$ and investigate the
migration law $\hat{y}(\hat{t})=y(t/\tau)/R_0$ for each value of $\nu$. Then we discuss the influence of $\nu$ and
$\lambda$.

\begin{figure}
\resizebox{\columnwidth}{!}{\includegraphics{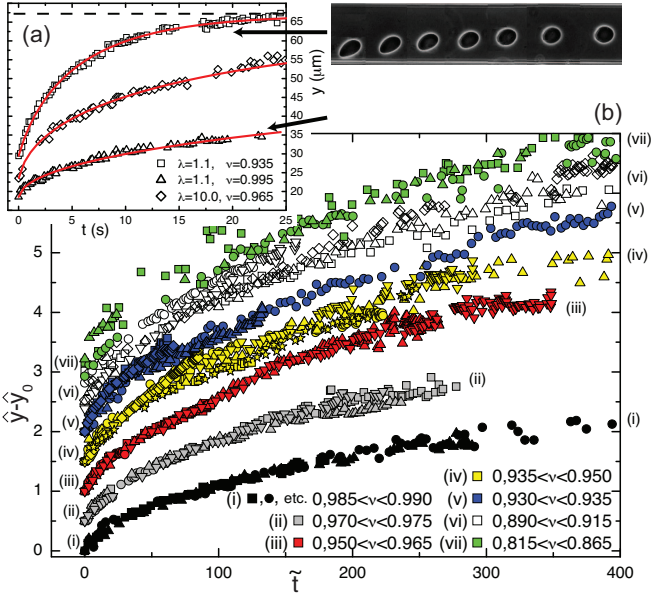}}
\caption{(color online) (a): Experimental time evolution of the
lateral position $y$ for 3 vesicles of similar size (19$\pm1\,\mu$m) and imposed flow velocity $v_0=920\,\mu$m.s$^{-1}$ but different $\nu$ and $\lambda$. The
dashed line indicates the center-line  (which will be reached much later by two of the vesicles). The solid line shows the
$y(t)$ curve obtained from the fit by the numerical solution of
Eq.~\ref{eq:eqmig}. (b): Evolution of $\hat{y}-\hat{y}_0$ versus
$\tilde{t}$ for vesicles with $\lambda=1.1$ and different $\nu$. For each $\nu$ interval, the different symbols correspond to different vesicles, whose $\hat{w}$ and $\hat{v}_0$ vary in the explored intervals, that are $1.8<\hat{w}<9.6$ and $400<\hat{v}_0<14800$. The corresponding curves clearly collapse on a single one. For clarity, the sets of curves corresponding to different $\nu$ intervals  are switched vertically with an increment of 0.5.} \label{fig:figrescaled}
\end{figure}

In the experiments, we used straight channels of height $
h_0=66.6\,\mu$m (in the direction of gravity $z$) and width $2 w$
(rectangular cross section) between 70 and 140 $\mu$m. The walls of the channels are made of
PDMS glued to a glass slide. The flow is induced by gravity, by connecting the inlet and the outlet to reservoirs at different heights. Vesicles are
prepared following the electroformation method. They are made of a
dioleoylphosphatidylcholine (DOPC) lipid bilayer enclosing an inner solution of sugar (sucrose or
glucose) in water or in a 1:4 glycerol-water (w:w) mixture. Samples
are diluted in a slightly hyperosmotic outer solution of the same
type, in order to deflate them by osmosis. Dextran can be added to
one of the solutions to modify the viscosity ratio $\lambda$. Vesicle size $R_0$ lies in the range 7-37$\,\mu$m while $v_0$ varies between 200 and $1100\,\mu$m.s$^{-1}$. Note that for our solutions of viscosity and density close to the one of water, the Reynolds number $R_e=\rho v_0 R_0/\eta_{out}$ is always lower than 4.10$^{-2}$.

A particular design of the upstream channel creates an initial
condition where incoming vesicles touch the $y=0$ wall in the
observation area and start to be lifted away from it. In particular,
they have already developed a nearly ellipsoidal shape tilted with
respect to the wall~\cite{cantat99,sukumaran01,abkarian02}.  The 2D fluid
velocity profile in the $xy$ plane where their center of mass lies is nearly
parabolic, since the rectangular cross section of the 3D channel
obeys $2w/h_0\le 3$~\cite{white}. Moreover, we wait for the flow to be established for a long time, resulting
in preliminary centering of the vesicles in the $z$ direction. The imposed velocity profile is thus written as
$v_x^{\infty}(\mathbf{r})=c \left(y w - {y^2}/{2}\right)$,  where $c
= 2 v_{0}/w^2$ is the curvature. A vesicle is tracked along its trajectory with a phase contrast
microscope, and the position $y$ of its center of mass is determined
by image processing, starting with  $y(t=0)=y_0$, where $y_0$
is the position just before lift-off, which is close to $R_0$.

The evolution with time of the $y$ position of two vesicles with $\lambda=1.1$ is shown in
Fig.~\ref{fig:figrescaled}(a). The vesicles quickly move away from
the wall, then the migration velocity decreases to zero as they
approach the center-line. Along their trajectory, they continuously
deform from a tilted ellipsoid to a symmetric bullet-like shape.
For a given $\nu$, the function $\hat{y}(\hat{t})$ depends a priori on both parameters
$\hat{w}$ and $\hat{v}_0$. In order to determine this functional dependence, which is not known {\itshape a priori}, we rescale the time variable. The choice of a relevant time scale is not
obvious, however. Indeed,  while the inverse of the shear rate
yields a natural scale, this is not an adequate choice since the
shear rate is not constant along the trajectory. The trick is  to
rescale each infinitesimal time step $dt$ around the time $t$ by the
local shear rate $\dot{\gamma}(y)=dv^{\infty}/dy=c(w-y)$ of the
unperturbed flow at the position $y(t)$. The new dimensionless
time-like parameter is then obtained by integrating the rescaled
time steps: $\tilde{t}=\int_0^t \dot{\gamma}(y) d t'= c \int_0^t \big[w-y(t')\big] d t'.$ 
$\tilde{t}$ accounts for the history of the shear rates experienced by
the vesicle along its trajectory. The raw data (not shown) for the
migration velocity spreads over more than a decade in the parameters
space. Interestingly, as shown in Fig.~\ref{fig:figrescaled}(b), all
experimental curves $\hat{y}(\tilde{t})$ for a given $\nu$ (or rather, a tiny interval around it)  collapse, whatever the values
of $\hat{w}$ and $\hat{v}_0$ within the explored range. A log-log plot
(not shown) of each master curve $\hat{y}(\tilde{t})-\hat{y}_0$ is
linear, a clear signature of a  power law behavior
$\hat{y}(\tilde{t})-\hat{y}_0=\beta \tilde{t}^{\alpha}$, where the
dimensionless parameters $\alpha$ and $\beta$ are  thus independent
from $\hat{w}$ and $\hat{v}_0$. The
lateral migration velocity $\hat{v}_m \equiv \hat{\dot{y}}$ as a function of the
position $\hat{y}$ and  $(\hat{w},\hat{v}_0)$ is then easily
extracted:\begin{equation} \hat{v}_m=\xi\frac{
\hat{\dot{\gamma}}(\hat{y})}{(\hat{y}-\hat{y}_0)^{\delta}}, \ \ \mbox{    that is:    }\ \ \dot{y}=\xi \frac{R_0^{\delta+1} \dot{\gamma}(y)}{(y-y_0)^{\delta}}.\label{eq:eqmig}
\end{equation}

This law constitutes the central result of our finding.
In the range $0.970<\nu<0.975$ we find for instance $\xi \equiv \alpha
\beta^{1/\alpha}= 1.2\times 10^{-2} \pm 0.2\times 10^{-2}$ and
$\delta\equiv 1/\alpha-1=1\pm 0.1$.  The error bars for these coefficients are mainly due to the uncertainties on the measure of $\nu$, of order $\pm 0.005$, since, as we shall see, the velocity depends on the reduced volume. Note that the differential equation~\ref{eq:eqmig} has no analytical solution but can be easily solved numerically and the result used to fit the raw data $y(t)$ without rescaling procedure (Fig.~\ref{fig:figrescaled}(a)).

\begin{figure}
\resizebox{\columnwidth}{!}{\includegraphics{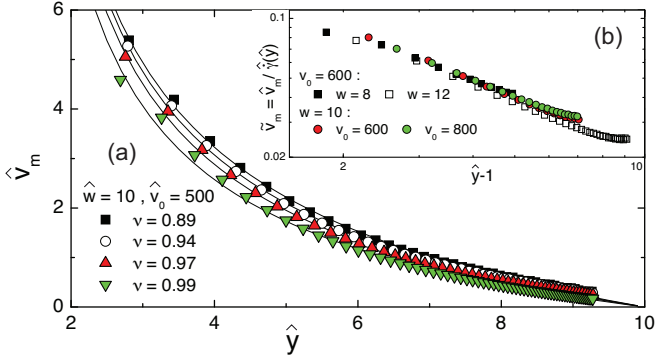}}
\caption{(color online) Simulations. (a): migration velocity $\hat{v}_m$ versus lateral position for different reduced volumes. Solid lines show the fits to Eq.~\ref{eq:eqmig}. (b): log-log plot of $\tilde{v}_m$ versus $\hat{y}-1$ for different $\hat{w}$ and $\hat{v}_0$ and $\nu=0.97$.} \label{fig:figsimul}
\end{figure}

In the simulations, we studied 2D neutrally buoyant vesicles (the 2D geometry captures the
essential features) having no viscosity contrast ($\lambda=1$).
The fluids flows inside and outside the vesicle are governed by the
Stokes equations ($\mathbf{r_m}\in\partial\Omega$  is a membrane
point):
\begin{equation}\label{eq:stokes}
-\nabla p(\mathbf{r}) + \eta \nabla^{2}\mathbf{v}(\mathbf{r})
=-\delta(\mathbf{r}-\mathbf{r_m})\mathbf{f}(\mathbf{r}),\,\, \nabla
\cdot \mathbf{v}(\mathbf{r}) = 0,
\end{equation}
where $p$ is the pressure, $\mathbf{v}$ is the velocity and
$\mathbf{f}$ the membrane force, given by Eq.~8 of
Ref.~\cite{kaoui08}. Thanks to the linearity of Eqs.
\ref{eq:stokes}, we solve them using a boundary integral method~\cite{pozrikidis9202} adapted to vesicle problems~\cite{sukumaran01,cantat03,beaucourt04,kaoui08}. The membrane
velocity is then given by the following integral equation, that we
solve numerically:
\begin{equation}
v_{i}(\mathbf{r_m})=\frac{1}{4\pi\eta}\oint_{\partial \Omega
}\!\!\mathbf{G}_{ij}^{W}(\mathbf{r_m},\mathbf{r}^{\prime
})f_{j}(\mathbf{r}^{\prime}) ds(\mathbf{r}^{\prime})+
v_{i}^{\infty}(\mathbf{r_m}), \label{eq: vel}
\end{equation}
where $\mathbf{G}_{ij}^{W}$ is the Green's function for a fluid
bounded by a steady infinite plane wall located at $y=0$:
\begin{eqnarray}
\mathbf{G}_{ij}^{W}(\mathbf{r},\mathbf{r}^{\prime})&=&
\mathbf{G}_{ij}(\mathbf{r}-\mathbf{r}^{\prime})-
\mathbf{G}_{ij}(\mathbf{r}-\mathbf{r}^{\prime}_{I}) \nonumber \\&+&
2r_y^{\prime
2}\mathbf{G}_{ij}^{D}(\mathbf{r}-\mathbf{r}^{\prime}_{I})-
2r_y^{\prime}\mathbf{G}_{ij}^{SD}(\mathbf{r}-\mathbf{r}^{\prime}_{I}).
\end{eqnarray}
$\mathbf{G}_{ij}(\mathbf{r})=-\delta_{ij}\text{ln}r+\frac{r_{i}r_{j}}{r^{2}}$
is the Green's function for an unbounded fluid, or
Stokeslet,
$\mathbf{r}^{\prime}_{I}=(r_{x}^{\prime},-r_{y}^{\prime})$ is the
image of $\mathbf{r}^{\prime}$ with respect to the wall. The
function
\begin{equation}
\mathbf{G}_{ij}^{D}(\mathbf{r})=(\delta_{jx}-\delta_{jy})\left(
\frac{\delta_{ij}}{r^{2}}-2 \frac{r_{i}r_{j}}{r^{2}} \right)
\end{equation}
is the Stokeslet doublet, and the source doublet is
\begin{equation}
\mathbf{G}_{ij}^{SD}(\mathbf{r})=r_{y}\mathbf{G}_{ij}^{D}(\mathbf{r})+
(\delta_{jx}-\delta_{jy})\frac{\delta_{jy}r_i-\delta_{iy}r_j}{r^{2}}.
\end{equation}
The evolution of the vesicle's shape and location are obtained by updating every
membrane point using a Euler scheme: $\mathbf{r_m}(t+\Delta
t)=\mathbf{v}(\mathbf{r_m},t)dt+\mathbf{r_m}(t)$.

Note that we only consider one wall at $y=0$. Provided $\hat{w}\ge 8$,
the vesicle reaches and stays on the $\hat{y}=\hat{w}$ line at long times,
even without the symmetric wall at $\hat{y}=2 \hat{w}$. This seems to indicate
that for $\hat{w}\ge 8$, migration forces due to the curvature
dominate over wall effects near the center-line.

\begin{figure}
\resizebox{\columnwidth}{!}{\includegraphics{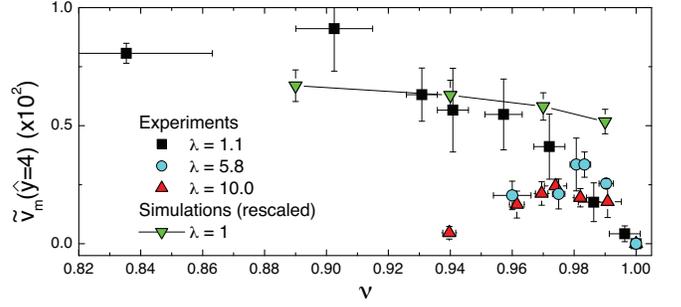}} \caption{(color
online) Reduced migration velocity $\tilde{v}_m$ of a vesicle at
position $\hat{y}=4$ versus its reduced volume. For readability, the simulations data are uniformly
rescaled by a factor $\sim 0.1$.} \label{fig:fignu}
\end{figure}
The variations of $\hat{v}_m$ with $\hat{y}$ are shown in
Fig.~\ref{fig:figsimul}(a) for four different reduced volumes and given width and flow velocity. They are well described by Eq.~\ref{eq:eqmig}. The adequation to this law is confirmed by considering varying $8<\hat{w}<12$ and $600<\hat{v}_0<800$. Following Eq.~\ref{eq:eqmig}, it is convenient to rescale the migration velocity in such a way that it should not depend either on $\hat{w}$, nor on $\hat{v}_0$:
$\tilde{v}_m\equiv
\hat{v}_m/\hat{\dot{\gamma}}(\hat{y})=\xi/(\hat{y}-\hat{y}_0)^{\delta}$. The variations of $\tilde{v}_m$  with $\hat{y}-1$ are shown in
Fig.~\ref{fig:figsimul}(b) for $\nu=0.97$. The collapse on a single line (in log-log scale) is in accordance with Eq.~\ref{eq:eqmig}. We find $\delta\simeq 0.8\pm0.1$ and $\xi\simeq 0.1$. The
agreement between experiments and simulations regarding the exponent
$\delta$ is quite satisfactory. However, numerical studies
overestimate the amplitude $\xi$. This is attributed to the 2D
character (actually a translationally invariant form in the $z$
direction), causing an enhancement of the lift force.  While 2D simulations have captured some interesting facts, a 3D simulation is necessary before drawing general conclusive answers, and we plan to investigate this problem in the future.

We now discuss the dependencies on  $\nu$ and $\lambda$; for all the values of $\nu$ and $\lambda$  explored here, the experimental and numerical curves  are still very well fitted by the law given by the resolution of Eq.~\ref{eq:eqmig} (see Fig.~\ref{fig:figrescaled}(a)).
Values for $\tilde{v}_m$ at $\hat{y}=4$ are reported on
Fig.~\ref{fig:fignu}. For $\lambda\simeq 1$, in the ranges $0.83\le\nu\le 1$ and $0.89\le\nu\le 0.99$ offered respectively by the experiments and the simulations, we find  an increasing migration velocity with decreasing $\nu$ (as seen on Fig.~\ref{fig:figsimul}(a)).

Results for higher $\lambda$ are available from the experiments. The new feature is the non monotonous behavior of $\tilde{v}_m$ as a function of $\nu$. When $\nu$ is decreased from 1, the migration velocity first increases (as for $\lambda=1.1$), then reaches a maximum and decreases back to a very low value (Fig.~\ref{fig:fignu}).

This non-monotonous behavior of the velocity can be understood on a
general physical basis. A spherical vesicle ($\nu=1$), does not
migrate owing to the fore-aft symmetry. On the other hand, when
$\nu$ decreases, the vesicle finally switches from a tank-treading
to a tumbling motion~\cite{mader06}. In the latter regime, no global
migration should occur either because the averaging over the different
orientations of the vesicle during one rotation period leads to an
almost symmetrical configuration. From these considerations we infer a maximal
velocity at a given value of $\nu$. Note that when $\lambda \simeq
1$, no tumbling motion occurs whatever $\nu$, so that a monotonous
evolution of the velocity with $\nu$ is observed. The evolution with
$\nu$ and $\lambda$ is in qualitative agreement with the
predictions made by Olla for the migration velocity of a (shape
preserving ellipsoidal) vesicle in the case of a simple shear flow
bounded by a  wall, although the scaling is different~\cite{olla97}.

In conclusion, our experiments and
simulations yield a similarity law for the lateral
migration velocity of a vesicle in a bounded Poiseuille flow as a
function of its distance to the walls and to the center-line, its
effective radius, the channel's width and the flow velocity. We 
showed that the effects of the walls and of the curvature of the
velocity field are coupled in a non linear manner: curvature not
only induces migration~\cite{kaoui08} but also affects the shape and
orientation, which affects the lift force. The law $v_m \sim
\dot{\gamma}(y)/y$ markedly differs from what the naive
extrapolation of the results for a vesicle near a wall and in a
linear shear flow would give: $v_m \sim \dot{\gamma}(y)/y^{2}$.

Deflating a spherical vesicle increases its deformability, thus its
asymmetry under shear, and leads to higher migration velocities.
However, beyond a given viscosity ratio, the tank-treading to
tumbling transition is approached when the deflation increases, and
the migration velocity undergoes a decline which can be understood
on the ground of general symmetry considerations. \\

Authors thank G. Danker and V. Vitkova for fruitful discussions, P. Ballet for technical assistance, and CNES and ANR
(MOSICOB) for financial support. Financial support from PAI
Volubilis (grant MA/06/144) is acknowledged. G.C. acknowledges a
fellowship from CNES. B.K. acknowledges PhD financial support from
CNRST (grant b4/015).

\end{document}